\documentclass{article}
\usepackage{epsfig}
\usepackage{latexsym}
    \newcommand{\ba}{\begin{eqnarray}}
    \newcommand{\ea}{\end{eqnarray}}
    \newcommand{\be}{\begin{equation}}
    \newcommand{\ee}{\end{equation}}

    \newcommand{\AmS}{{\protect\the\textfont2
      A\kern-.1667em\lower.5ex\hbox{M}\kern-.125emS}}

\DeclareFontFamily{OT1}{rsfs}{}
\DeclareFontShape{OT1}{rsfs}{m}{n}{ <-7> rsfs5 <7-10> rsfs7 <10->
rsfs10}{} \DeclareMathAlphabet{\mycal}{OT1}{rsfs}{m}{n}
\def\scri{{\mycal I}}

\begin{document}
\title{Hawking radiation from spherically symmetrical gravitational collapse to an extremal R-N black hole for a charged scalar field}
\author{\begin{tabular}{c}\bigskip Hongbao Zhang$^{1}$\footnote{Email: hongbaozhang@pku.edu.cn}, Zhoujian Cao$^{2}$, Chongshou Gao$^{1, 3}$\\  \smallskip$^1$School of Physics, Peking University, Beijing, 100871, PRC
\\ \smallskip$^2$Department of Physics, Beijing Normal University, Beijing,
100875, PRC\\$^3$Institute of Theoretical Physics, Academia
Sinica, Beijing, 100080, PRC\end{tabular}} \maketitle
\begin{abstract}
Sijie Gao has recently investigated Hawking radiation from
spherically symmetrical gravitational collapse to an extremal R-N
black hole for a real scalar field. Especially he estimated the
upper bound for the expected number of particles in any wave
packet belonging to $\mathcal{H}_{out}$ spontaneously produced
from the state $|0\rangle_{in}$, which confirms the traditional
belief that extremal black holes do not radiate particles. Making
some modifications, we demonstrate that the analysis can go
through for a charged scalar field.
\end{abstract}
\section{Introduction}
Particle creation by black holes (Hawking radiation) has been
widely studied since 1970s \cite{Hawking,Wald75}. One of the most
important results is that a black hole radiates particles with the
same spectrum as a black body with temperature
$T=\frac{\hbar\kappa}{2\pi k}$, where $\kappa$ is the surface
gravity of the black hole and $k$ is Boltzmann's constant.

Although the standard Hawking radiation is only derived for a
spacetime appropriate to a collapsing body which settles down to a
non-extremal black hole at the late stage, it is generally
accepted that an extremal black hole has zero temperature (zero
surface gravity) and consequently no particles are produced (not
for superradiance modes if exist). However Liberati \emph{et al}.
\cite{Liberati} pointed out that the generalization to extremal
black holes from non-extremal black holes is not trivial. One
important point is that Kruskal transformation for non-extremal
black holes, which plays a crucial role in calculating the
particle creation, need be modified for extremal black holes.

The modified Kruskal transformation given in \cite{Liberati} is
briefly reviewed as follows. Start with the usual form of the R-N
geometry with parameters $M$ and $Q$\footnote{Without loss of
generalization, only discuss the case of $Q>0$.}
\begin{equation}
ds^{2}=\frac{\Delta}{r^{2}}dt^{2}-\frac{r^{2}}{\Delta}dr^{2}-r^{2}(d\theta^{2}+\sin^{2}\theta
d\varphi^{2})
\end{equation}
where $\Delta=r^{2}-2rM+Q^{2}$. In the extremal case $Q=M$, the
tortoise coordinate $r_{*}$ is given by
\begin{equation}
r_{*}=r+2M[\ln(r-M)-\frac{M}{2(r-M)}]
\end{equation}

Define the advanced time $v$ and retarded time $u$ as
\begin{eqnarray}
v&=&t+r_{*}\nonumber\\
u&=&t-r_{*}
\end{eqnarray}
the modified Kruskal tranformation for extremal R-N black holes is
[3]
\begin{eqnarray}
u&=&-4M[\ln(-U)+\frac{M}{2U}]\nonumber\\
v&=&4M(\ln V-\frac{M}{2V})
\end{eqnarray}
where $U$ and $V$ are regular across the past and future horizons
of the extended spacetime.

Using this modified Kruskal transformation, Liberati \emph{et al}.
\cite{Liberati} find that the extremal black hole does not behave
as a thermal object and can not be regarded as the thermodynamic
limit of a non-extremal black hole. However, Gao \cite{Gao}
pointed out some deficiencies in the analysis of \cite{Liberati}
and circumvented them by focusing on the wave packet solutions
with unit norm. The result in \cite{Gao} confirms that extremal
black holes do not radiate particles.

Nevertheless, \cite{Liberati,Gao} only involves the calculation
for the massless real scalar field. this paper will focus on the
massless charged scalar field in an extremal R-N black hole formed
from a collapsing spherically symmetrical star, which makes two
major differences from the massless real scalar field: One is that
not only particle creation but also antiparticle creation need be
calculated; the other is that the influence from the
electromagnetic field need be considered. Our calculation follows
the similar steps to \cite{Gao}.
\section{Calculation of particle creation}
\subsection{Construction of the wave packet at future null infinity}
Our purpose in this subsection is to construct the wave packet
with unit norm at future null infinity $\scri^+$. Start with the
wave equation in the appendix with the massless and zero potential
case\footnote{Without loss of generalization, let $q>0$.}
\begin{equation}
\frac{1}{\sqrt{-g}}(\partial_{i}+iqA_{i})[\sqrt{-g}g^{ij}(\partial_{j}+iqA_{j})]\phi=0
\end{equation}

In the region outside the collapsing star, the spacetime is
described by the extremal R-N metric. Let
$\phi=\frac{1}{r}f(t,r)Y_{lm}(\theta,\varphi)$, where
$Y_{lm}(\theta,\varphi)$ is spherical harmonic. Thus if choose
$A_{a}=-\frac{Q}{r}(dt)_{a}$, then outside the collapsing star
\begin{equation}
\frac{\partial^{2}}{\partial^{2}t}f-\frac{\partial^{2}}{\partial^{2}r_{*}}f-[\frac{2iqQ}{r}\frac{\partial}{\partial
t}f+\frac{(qQ)^{2}}{r^{2}}f]+V(r)f=0
\end{equation}
where
\begin{equation}
V(r)=\frac{(M-r)^{2}}{r^{6}}[2M(M-r)+l(l+1)r^{2}]
\end{equation}

It is easy to know
\begin{equation}
F_{w_{0}lm}=\frac{1}{r}e^{i\omega_{0}u}Y_{lm}(\theta,\varphi)\label{8}
\end{equation}
is a solution near $\scri^+$. Then following the procedure in
\cite{Gao}, construct the wave packet with frequency around
$\omega_{0}$ and centered on the retarded time $u=2\pi n$ as
\begin{equation}
P_{n\omega_{0}lm}=\frac{1}{r}z_{n\omega_{0}}Y_{lm}(\theta,\varphi)\label{9}
\end{equation}
 where
\begin{equation}
z_{n\omega_{0}}=\frac{N}{\sqrt{2\pi}}\int_{-\infty}^{\infty}g(\frac{\omega-\omega_{0}}{\epsilon})e^{i\omega(u-2\pi
n)}d\omega
\end{equation}
where N is normalization constant and $g(x)$ is real $C^{\infty}$
function with compact support in $[-1,1]$. To guarantee
$P_{n\omega_{0}lm}$ has frequencies with the same sign as
$F_{\omega_{0}lm}$, require $0<\epsilon\ll|\omega_{0}| $. If let
$\frac{\omega-\omega_{0}}{\epsilon}=\hat{\omega}, u-2\pi
n=\hat{u}$, then
\begin{equation}
z_{n\omega_{0}}(u)=N\epsilon
e^{i\omega_{0}\hat{u}}\hat{g}(\epsilon\hat{u})\label{11}
\end{equation}
where
\begin{equation}
\hat{g}(x)=\frac{1}{\sqrt{2\pi}}\int_{-\infty}^{\infty}g(\hat{\omega})e^{i\hat{\omega}x}d\hat{\omega}
\end{equation}

The normalization constant N can be determined by the inner
product in the appendix
\begin{equation}
\langle\Omega^{-1}P,\Omega^{-1}P\rangle=(P,P)=1
\end{equation}

It is convenient to make the integral on $\scri^+$, and
accordingly
\begin{equation}
sign(-\omega_{0})i\int_{-\infty}^{\infty}\bar{z}(u)z'(u)-z(u)\bar{z}'(u)du=1
\end{equation}

Then the straightforward calculation gives \cite{Gao}
\begin{equation}
N=\frac{1}{\sqrt{sign(\omega_{0})(\beta\epsilon\omega_{0}+\gamma\epsilon^{2})}}
\end{equation}
where
\begin{eqnarray}
\beta&=&2\int_{-\infty}^{\infty}|g(\hat{\omega})|^{2}d\hat{\omega}\nonumber\\
\gamma&=&2\int_{-\infty}^{\infty}\hat{\omega}|g(\hat{\omega})|^{2}d\hat{\omega}
\end{eqnarray}

Thus (\ref{11}) becomes
\begin{equation}
z(u)=\frac{\sqrt{\epsilon}}{\sqrt{sign(\omega_{0})(\beta\omega_{0}+\gamma\epsilon)}}e^{i\omega_{0}\hat{u}}\hat{g}(\epsilon\hat{u})
\end{equation}
\subsection{Geometrical optics approximation}
To calculate the particle production at late times, the wave
packet (\ref{9}) need be propagated backwards from $\scri^+$ to
the past null infinity $\scri^-$. For simplicity, first
investigate the propagation of wave (\ref{8}); then the
propagation of (\ref{9}) can be got by superposition. A part of
wave (\ref{8}) will be scattered by the gravitational field and
electromagnetic field outside the collapsing star and will end up
on $\scri^-$ with the same frequency, which indicates this part
will not contribute the particle production. So we will only care
the remaining part which will propagate through the collapsing
star, eventually emerging to $\scri^-$.

Note geometrical optics approximation holds for the propagation of
the remaining part from near the future event horizon to $\scri^-$
\cite{Hawking}. Thus if set $v=0$ for the points where the null
curves generating the future event horizon intersects $\scri^-$;
for $-v$ positive and small
\begin{equation}
U=\alpha v
\end{equation}
Therefore near $\scri^-$, the remaining part takes the form as
\cite{Hawking}
\begin{equation}
F_{rem}=\left\{\begin{array}{r@{\quad\quad}l}
t(\omega_{0})\frac{1}{r}e^{iS(v)}Y_{lm}(\theta,\varphi)&v<0\\0&v>0
\end{array}\right.
\end{equation}
where $t(\omega_{0})$ is the transmission amplitude and
\begin{equation}
S(v)=(\omega_{0}-\frac{qQ}{M})\{-4M[\ln(-\alpha v
)+\frac{M}{2\alpha v}]\}
\end{equation}

Next assume that $t(\omega)$ varies negligibly over the frequency
interval $2\epsilon$, thus near $\scri^-$
\begin{equation}
P_{rem}=\left\{\begin{array}{r@{\quad\quad}l}
t(\omega_{0})\frac{1}{r}z(v)Y_{lm}(\theta,\varphi)&v<0\\0&v>0
\end{array}\right.
\end{equation}
where
\begin{equation}
z(v)=\frac{\sqrt{\epsilon}}{\sqrt{sign(\omega_{0})(\beta\omega_{0}+\gamma\epsilon)}}e^{i(\omega_{0}-\frac{qQ}{M})\hat{u}(v)}\hat{g}[\epsilon\hat{u}(v)]
\end{equation}
and
\begin{equation}
\hat{u}(v)=-4M\ln(-\alpha v)-\frac{2M^{2}}{\alpha v}-2n\pi
\end{equation}
\subsection{Calculation of particle creation}
According to the formulae in the appendix, the expected number of
particles spontaneously produced in the state represented by
$P_{n\omega_{0}lm}$ is given by
\begin{equation}
N_{n\omega_{0}lm}=2|t(\omega_{0})|^{2}\int_{0}^{\infty}\omega'|\hat{z}[sign(\omega_{0})\omega']|^{2}d\omega'
\end{equation}
where
\begin{eqnarray}
\hat{z}(\omega')&=&\frac{1}{\sqrt{2\pi}}\int_{-\infty}^{0}z(v)e^{i\omega'v}dv\nonumber\\
&=&\frac{\sqrt{\epsilon}}{\sqrt{sign(\omega_{0})2\pi
(\beta\omega_{0}+\gamma\epsilon)}}Z(\omega')
\end{eqnarray}
where
\begin{equation}
Z(\omega')=\int_{-\infty}^{0}e^{i\omega'_{0}\hat{u}(v)}e^{i\omega'v}\hat{g}[\epsilon\hat{u}(v)]dv
\end{equation}
where
\begin{equation}
\omega'_{0}=\omega_{0}-\frac{qQ}{M}
\end{equation}

In addition, by a simple rescaling, It is easy to know that
$N_{n\omega_{0}lm}$ is independent of the choice of $\alpha$ in
$\hat{u}(v)$. So without loss of generality, from now on let
\begin{equation}
\hat{u}(v)=-4M\ln(-v)-\frac{2M^{2}}{v}-2n\pi
\end{equation}

Since the traditional belief that extremal black holes do not
radiate particles is only for non-superradiance modes, next the
creation of particles will be calculated in the following two
cases: $(a) \omega_{0}>\frac{qQ}{M},\omega'>0$ \footnote{Require
$0<\epsilon\ll\omega'_{0}$ to guarantee $P_{n\omega_{0}lm}$ only
includes non-superradiance modes} or $(b) \omega_{0}<0,\omega'<0$.
The calculation here follows Gao's \cite{Gao} by some simple
modifications.
\begin{figure}
\centering
\scalebox{0.8}[0.8]{\includegraphics*[47pt,395pt][499pt,595pt]{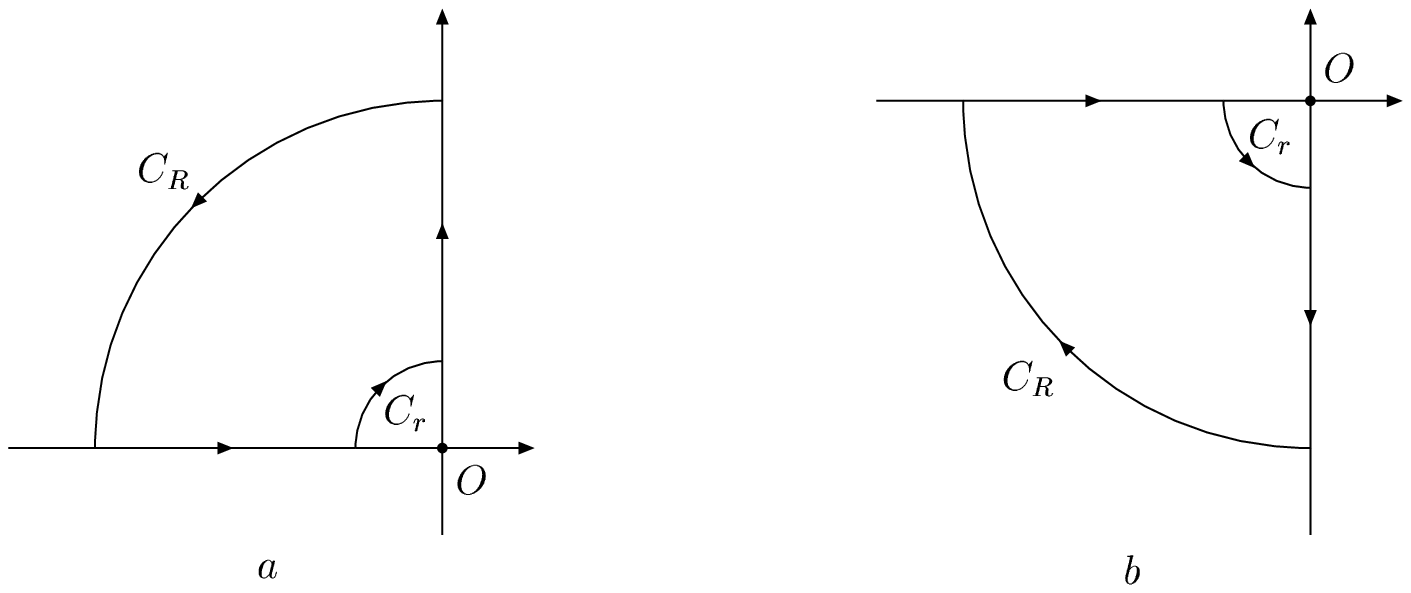}}
\caption[*]{\small The Wick rotation $(a)$ for
$\omega_{0}>\frac{qQ}{M},\omega'>0$ and $(b)$ for
$\omega_{0}<0,\omega'<0$ }\label{2}
\end{figure}

According to the theorem and the lemma in the appendix, make a
Wick rotation (See Figure \ref{2}). Thus if let $x=-i\omega' v$,
then
\begin{equation}
Z(\omega')=\frac{e^{-i\omega'_{0}2n\pi}}{i\omega'}\int_{0}^{\infty}e^{-x}e^{-i\omega'_{0}4M\ln(\frac{x}{i\omega'})}e^{-\frac{2M^{2}\omega'_{0}\omega'}{x}}\hat{g}[\epsilon\hat{u}(x)]dx
\end{equation}
where
\begin{equation}
\hat{u}(x)=-4M\ln(\frac{x}{i\omega'})+\frac{i2M^{2}\omega'}{x}-2n\pi
\end{equation}

Using the fact
$Im[\hat{u}(x)]=sign(\omega_{0})2M\pi+\frac{2M^{2}\omega'}{x}$ and
$|e^{-i\omega'_{0}4M\ln(\frac{x}{i\omega'})}|=e^{sign(-\omega_{0})2M\pi\omega'_{0}}$
when $x$ is real
\begin{eqnarray}
|Z(\omega')|&\leq&\frac{C_{k}e^{sign(-\omega_{0})2M\pi(\omega'_{0}-\epsilon)}}{sign(\omega_{0})\omega'}\int_{0}^{\infty}\frac{e^{-x}e^{-\frac{2M^{2}\omega'(\omega'_{0}-\epsilon)}{x}}}{[1+|\epsilon\hat{u}(x)|]^{k}}dx
\nonumber \\
&\leq&\frac{C_{k}e^{sign(-\omega_{0})2M\pi(\omega'_{0}-\epsilon)}}{sign(\omega_{0})\omega'}\int_{0}^{\infty}\frac{e^{-x}e^{-\frac{2M^{2}\omega'(\omega'_{0}-\epsilon)}{x}}}{|\epsilon\hat{u}(x)|^{k}}dx\label{31}
\end{eqnarray}

To proceed, first give a lower bound for $|\hat{u}(x)|$. Start
with
\begin{equation}
|\hat{u}(x)|^{2}=(2n\pi+4M\ln y)^{2}+(2M\pi+\frac{2M^{2}}{y})^{2}
\end{equation}
where $y=\frac{x}{sign(\omega_{0})\omega'}$. Thus it is not
difficult to find
\begin{equation}
|\hat{u}(x)|\geq c_{1}2n\pi
\end{equation}
where $c_{1}$ is a positive constant. Therefore
\begin{eqnarray}
|Z(\omega')|&\leq&\frac{C_{k}e^{sign(-\omega_{0})2M\pi(\omega'_{0}-\epsilon)}}{sign(\omega_{0})\omega'(\epsilon
c_{1}
)^{k}(2n\pi)^{k}}\int_{0}^{\infty}e^{-x}e^{-\frac{2M^{2}\omega'(\omega'_{0}-\epsilon)}{x}}dx \nonumber\\
&=&\frac{C_{k}e^{sign(-\omega_{0})2M\pi(\omega'_{0}-\epsilon)}2\sqrt{2M^{2}(\omega'_{0}-\epsilon)\omega'}K_{1}[2\sqrt{2M^{2}(\omega'_{0}-\epsilon)\omega'}]}{sign(\omega_{0})\omega'(\epsilon
c_{1} )^{k}(2n\pi)^{k}}\nonumber\\
\end{eqnarray}
where $K_{1}(z)$ is modified Bessel function \cite{WG}.

In order to investigate the bound in (\ref{31}) for
$sign(\omega_{0})\omega'\leq\Omega\ll 1$ more carefully, denote by
$G(\omega')$ the integral in (\ref{31}) for
$sign(\omega_{0})\omega'\leq\Omega\ll 1$ , i.e.
\begin{equation}
G(\omega')=\int_{0}^{\infty}\frac{e^{-x}e^{-\frac{2M^{2}\omega'(\omega'_{0}-\epsilon)}{x}}}{|\epsilon\hat{u}(x)|^{k}}dx
\end{equation}
and rewrite $|\hat{u}(x)|^{2}$ as
\begin{equation}
|\hat{u}(x)|^{2}=[F_{n}(\omega')+4M\ln
x]^{2}+[2M\pi+\frac{sign(\omega_{0})2M^{2}\omega'}{x}]^{2}
\end{equation}
where
\begin{equation}
F_{n}(\omega')=2n\pi-4M\ln [sign(\omega_{0})\omega']
\end{equation}

In addition, in the following discussion, assume $a\ll 1$ but
$an\gg 1$.

(a) $D_{1}=\{e^{-\frac{aF_{n}(\omega')}{4M}}\leq x \leq
e^{\frac{aF_{n}(\omega')}{4M}}$ and $
x\geq\frac{sign(\omega_{0})2M^{2}\omega'}{aF_{n}(\omega')}\}$
\begin{eqnarray}
G(\omega',D_{1})&\leq&\frac{1}{[\epsilon c_{2}F_{n}(\omega')]^{k}}\int_{0}^{\infty}e^{-x}e^{-\frac{2M^{2}\omega'(\omega'_{0}-\epsilon)}{x}}dx\nonumber\\
&=&\frac{2\sqrt{2M^{2}(\omega'_{0}-\epsilon)\omega'}K_{1}[2\sqrt{2M^{2}(\omega'_{0}-\epsilon)\omega'}]}{[\epsilon
c_{2} F_{n}(\omega')]^{k}}
\end{eqnarray}
where $c_{2}$ is a positive constant. For $z\rightarrow 0$,
$K_{1}(z)\sim\frac{1}{z}$ \cite{Guo}. Therefore
\begin{equation}
G(\omega',D_{1})\leq\frac{1}{[\epsilon c_{2}F_{n}(\omega')]^{k}}
\end{equation}
(b) $D_{2}=\{x\leq
\frac{sign(\omega_{0})2M^{2}\omega'}{aF_{n}(\omega')}\}$
\begin{equation}
G(\omega',D_{2})\leq
\frac{sign(\omega_{0})2M^{2}\omega'e^{sign(-\omega_{0})aF_{n}(\omega')(\omega'_{0}-\epsilon)}}{aF_{n}(\omega')(\epsilon
c_{1})^{k}(2n\pi)^{k}}
\end{equation}
(c) $D_{3}=\{x\in (0,\infty)|x\in\!\!\!\!\!/D_{1}\cup D_{2}\}$
\begin{eqnarray}
G(\omega',D_{3})&\leq& \int_{0}^{e^{-\frac{aF_{n}(\omega')}{4M}}}\frac{e^{-x}e^{-\frac{2M^{2}\omega'(\omega'_{0}-\epsilon)}{x}}}{|\epsilon\hat{u}(x)|^{k}}dx+\int_{e^{\frac{aF_{n}(\omega')}{4M}}}^{\infty}\frac{e^{-x}e^{-\frac{2M^{2}\omega'(\omega'_{0}-\epsilon)}{x}}}{|\epsilon\hat{u}(x)|^{k}}dx \nonumber\\
&\leq&\frac{e^{-\frac{aF_{n}(\omega')}{4M}}}{(\epsilon c_{1}
)^{k}(2n\pi)^{k}}+\frac{e^{-e^{\frac{aF_{n}(\omega')}{4M}}}}{(\epsilon
c_{1} )^{k}(2n\pi)^{k}}
\end{eqnarray}
Thus
\begin{equation}
G(\omega')\leq \frac{1}{[\epsilon
c_{3}(\omega_{0})F_{n}(\omega')]^{k}}
\end{equation}
where $c_{3}(\omega_{0})$ is a positive constant.

Thus
\begin{equation}
|\hat{z}[sign(\omega_{0})\omega'\geq\Omega]|\leq\frac{\sqrt{\epsilon}C_{k}e^{sign(-\omega_{0})2M\pi(\omega'_{0}-\epsilon)}}{\sqrt{sign(\omega_{0})2\pi
(\beta\omega_{0}+\gamma\epsilon)} \omega'[\epsilon
c_{3}(\omega_{0})F_{n}(\omega')]^{k}}
\end{equation}
and
\begin{eqnarray}
|\hat{z}[sign(\omega_{0})\omega'>\Omega]|\leq
\frac{\sqrt{\epsilon}C_{k}e^{sign(-\omega_{0})2M\pi(\omega'_{0}-\epsilon)}2\sqrt{2M^{2}(\omega'_{0}-\epsilon)\omega'}K_{1}[2\sqrt{2M^{2}(\omega'_{0}-\epsilon)\omega'}]}{sign(\omega_{0})\sqrt{sign(\omega_{0})2\pi
(\beta\omega_{0}+\gamma\epsilon)} \omega'(\epsilon c_{1}
)^{k}(2n\pi)^{k}}\nonumber
\end{eqnarray}
\begin{equation}
\end{equation}
Therefore
\begin{eqnarray}
N_{n\omega_{0}lm}&\leq&2|t(\omega_{0})|^{2}\{\frac{\epsilon
C_{k}^{2}e^{sign(-\omega_{0})4M\pi(\omega'_{0}-\epsilon)}}{sign(\omega_{0})2\pi
(\beta\omega_{0}+\gamma\epsilon)(\epsilon
c_{3})^{2k}}\int_{0}^{\Omega}\frac{1}{\omega'(2n\pi-4M\ln
\omega')^{2k}}d\omega'\nonumber \\&+&\frac{8\epsilon
C_{k}^{2}e^{sign(-\omega_{0})4M\pi(\omega'_{0}-\epsilon)}M^{2}(\omega'_{0}-\epsilon)}{sign(\omega_{0})2\pi
(\beta\omega_{0}+\gamma\epsilon)(\epsilon
c_{1})^{2k}(2n\pi)^{2k}}\int_{\Omega}^{\infty}\frac{K_{1}^{2}[2\sqrt{2M^{2}(\omega'_{0}-\epsilon)\omega'}]}{\omega'}d\omega'\}\nonumber
\end{eqnarray}
\begin{equation}
\end{equation}
For $z\rightarrow\infty$,
$k_{1}(z)\sim\sqrt{\frac{\pi}{2z}}e^{-z}$ \cite{Guo}. Thus
\begin{equation}
N_{n\omega_{0}lm}\leq\frac{2c_{4}(\omega_{0})|t(\omega_{0})|^{2}}{(2n\pi)^{2k-1}}\label{46}
\end{equation}
where $c_{4}(\omega_{0})$ is a positive constant.
\section{Conclusion and discussion}
Since $2n\pi$ represents time, (\ref{46}) shows that for any
non-superradiance mode associated with a wave packet, the rate of
particle and antiparticle creation drops off to zero faster than
any inverse power of time at late times. This result confirms that
extremal black holes do not radiate particles as \cite{Gao}. In
addition, it is obvious that our result is immediately applicable
to the case of a charged scalar field in an extremal K-N black
hole; since the modified Kruskal transformation here is exactly
the same as the extremal R-N case \cite{Rothman}.

\section*{Acknowledgments}
We would like to give our thanks to Dr. Sijie Gao for many helpful
discussions throughout the whole project. In addition, this work
was supported in part by the National Nature Science of China
(90103019) and the Doctoral Programme Foundation of Institute of
Higher Education, the State Education Commission of China
(2000000147).

\section*{Appendices}
\begin{appendix}\setcounter{equation}{0}
\subsection*{A  The charged (complex) scalar quantum field in
curved spacetimes} Here the charged scalar quantum field in curved
spacetimes is briefly reworked \cite{Wald79}. Start with the
gauge-invariant\footnote{Under the transformations:
$\phi'=e^{-iq\tau}\phi,
 A'_{a}=A_{a}+\bigtriangledown_{a}\tau$} wave equation
\begin{equation}
(\bigtriangledown_{a}+iqA_{a})(\bigtriangledown^{a}+iqA^{a})\phi-\frac{1}{6}R\phi+m^{2}\phi+V\phi=0
\end{equation}

The Klein-Gordon inner product
\begin{equation}
(\phi,\psi)=i\int_{\Sigma}[\bar{\phi}(\bigtriangledown_{a}+iqA_{a})\psi-\psi(\bigtriangledown_{a}-iqA_{a})\bar{\phi}]n^{a}dV
\end{equation}
is gauge-invariant and independent of choice of Cauchy surface
$\Sigma$ .

For the free field, we define the one-particle Hilbert space
$\mathcal{H}$ as
\begin{equation}
\mathcal{H}=\mathcal{H}^{+}\bigoplus\bar{\mathcal{H}}^{-}
\end{equation}
where $\mathcal{H}^{+}$ is the Hilbert space of positive frequency
solutions of the Klein-Gordon equation with the above inner
product; $\mathcal{H}^{-}$ is the Hilbert space of negative
frequency solutions with the inner product given by minus the
above Klein-Gordon inner product, and the bar denotes the complex
conjugate (dual) operation.

The Hilbert space of free-field states is taken to be
$\mathcal{F}_{s}(\mathcal{H})$.Annihilation and creation operators
are be defined as usual. Then the field operator $\hat{\phi}$ is
defined as
\begin{equation}
\hat{\phi}=\varrho_{i}a(\bar{\varrho}^{i})+c^{\dagger}(\lambda_{i})\bar{\lambda}^{i}
\end{equation}
where $\{\varrho_{i}\}$, $\{\lambda_{i}\}$ are respectively the
orthonormal bases of $\mathcal{H}^{+}$, $\bar{\mathcal{H}}^{-}$.

Then according to the usual procedure, the scattering matrix can
be constructed. Here only give the expected number of particles in
states belonging to $\mathcal{H}_{out}$ spontaneously produced
from the state $|0\rangle_{in}$
\begin{eqnarray}
\langle
\hat{N}(\varrho_{i})\rangle_{out}&=&|\overline{B\varrho_{i}}|_{in}^{2}\nonumber
\\\langle \hat{N}(\lambda_{i})\rangle_{out}&=&|A\bar{\lambda}_{i}|_{in}^{2}
\end{eqnarray}
where $B$ denotes taking the past negative part of solutions which
are purely positive frequency in the future, and $A$ denotes
taking the past positive part of solutions which are purely
negative frequency in the future.

Now consider some conformally invariant properties for the charged
scalar field. Make the conformal transformation
$\tilde{g}_{ab}=\Omega^{2}g_{ab}$, then
\begin{eqnarray}
&&[\tilde{g}^{ab}(\tilde{\bigtriangledown}_{a}+iqA_{a})(\tilde{\bigtriangledown}_{b}+iqA_{b})-\frac{1}{6}\tilde{R}+\Omega^{-2}(m^{2}+V)](\Omega^{-1}\phi)=\nonumber\\
&&\Omega^{-3}[g^{ab}(\bigtriangledown_{a}+iqA_{a})(\bigtriangledown_{b}+iqA_{b})-\frac{1}{6}R+m^{2}+V]\phi
\end{eqnarray}

Thus the conformal weight for the charged scalar field is -1.
Furthermore
\begin{equation}
\langle\Omega^{-1}\phi,\Omega^{-1}\psi\rangle=(\phi,\psi)
\end{equation}

In order to consider the null infinity $\scri^-$ of R-N black
hole, select the conformal factor $\Omega=\frac{1}{r}$. Thus
\begin{eqnarray}
d\tilde{s}^{2}&=&\Omega^{4}\Delta
dv^{2}+2dvd\Omega-(d\theta^{2}+\sin^{2}\theta
d\varphi^{2})\nonumber\\
d\tilde{s}^{2}&=&\Omega^{4}\Delta
du^{2}-2dud\Omega-(d\theta^{2}+\sin^{2}\theta d\varphi^{2})
\end{eqnarray}

Obviously, the above metrics are non-degenerate at $\Omega=0$.
Furthermore, $\scri^-$ can be defined as $\Omega=0,
-\infty<v<\infty$; $\scri^+$ as $\Omega=0, -\infty<u<\infty$. In
addition, if the Cauchy surface chosen includes either $\scri^-$
or $\scri^+$, choose \cite{Wald84}
\begin{eqnarray}
n^{a}=(\frac{\partial}{\partial v})^{a},&&dV=\sin\theta dvd\theta
d\varphi
\end{eqnarray}
for $\scri^-$;
\begin{eqnarray}
n^{a}=(\frac{\partial}{\partial u})^{a},&&dV=\sin\theta dud\theta
d\varphi
\end{eqnarray}
for $\scri^+$.
\subsection*{B One theorem and one
lemma}\label{B}Here the theorem and lemma useful to our
calculation of the creation of particles are briefly reviewed.

\textbf{Theorem} \cite{Gao,RS} Let $g:R\rightarrow R$ be a
$C^{\infty}$ function with compact support in $[-1,1 ]$. Then, the
Fourier transform, $\hat{g}(\zeta)$ is an entire analytic function
of $\zeta$ such that for all $k>0$
\begin{equation}
|\hat{g}(\zeta)|\leq\frac{C_{k}e^{|Im\zeta|}}{(1+|\zeta|)^{k}}
\end{equation}
for all $\zeta\in C$, where $C_{k}>0$ is a constant which depends
on $k$ and $g$.

\textbf{Lemma} \cite{Gao,Guo} If $F(z)$ satisfies
$\lim_{|z|\rightarrow \infty}|F(z)|=0$ in the second (third)
quadrant, then
\begin{equation}
\lim_{R\rightarrow\infty}\int_{C_{R}}e^{icz}F(z)=0
\end{equation}
where $c>0 (c<0)$ is a constant.
\end{appendix}

\end{document}